\documentclass[aps,prb,twocolumn,superscriptaddress,showpacs]{revtex4-1}
\usepackage{graphicx}
\usepackage{color}
\usepackage{amsmath}

\begin{document}

\preprint{}

\title{Graphite under uniaxial compression along c-axis: a new parameter relates out-of-plane strain to in-plane phonon frequency}


\author{Y. W. Sun}
\email[corresponding author: ]{yiwei.sun@qmul.ac.uk}
\affiliation{School of Physics and Astronomy, Queen Mary University of London, London E1 4NS, UK}
\author{D. Holec}
\email[more calculation details: ]{david.holec@unileoben.ac.at}
\affiliation{Department of Physical Metallurgy and Materials Testing, Montanuniversit\"{a}t Leoben, 8700 Leoben, Austria}
\author{D. J. Dunstan}
\affiliation{School of Physics and Astronomy, Queen Mary University of London, London E1 4NS, UK}

\date{\today}

\begin{abstract}
Stacking graphene sheets forms graphite. Two in-plane vibrational modes of graphite, E$_{1u}$ and E$_{2g}^{(2)}$, are derived from the graphene E$_{2g}$ mode, the shifts of which under compression are considered as results of the in-plane bond shortening. Values of the Gr\"{u}neisen parameter have been reported to quantify such relation. However, the reason why the shift rates of these three modes with pressure differ is unclear. In this work, we introduce \textcolor{blue}{new parameters $\gamma\prime_{E_{2g}}$=-0.0131 and $\gamma\prime_{E_{1u}}$=0.0585} to quantify the contribution of out-of-plane strain to the shift of the in-plane vibrational frequencies, suggesting that the compression of the $\pi$-electrons plays a non-negligible part in both graphite and graphene under high pressure. 
\end{abstract}


\maketitle


\section{Introduction}
Graphene has been investigated intensively since its discovery in 2004 \cite{Novoselov2004}, for its unique mechanical and electrical properties \cite{Novoselov2005,Zhang2005}. The motivation to study graphene under strain has been illustrated in the literature \cite{Proctor2009,Mohiuddin2009,Huang2009,Ni2008,Mohr2009}. In brief, strain can modify the properties of graphene to meet specific scientific or technological needs. Therefore, strain determination and monitoring are of critical application importance and contribute to the characterization and understanding of this remarkable material. Strain is related to phonon frequencies, which can be directly obtained by experiments.  The analysis, rather naturally, is two-dimensional. Thus the frequency shifts of the graphene E$_{2g}$ mode are considered as entirely induced by in-plane strain.

Stacking graphene sheets forms graphite. The in-plane vibrational modes E$_{1u}$ and E$_{2g}^{(2)}$ of graphite are derived from the graphene E$_{2g}$ mode, and from the two dimensional analysis, all these three modes were expected to have the same shift rate \cite{Thomsen2002,Mohiuddin2009,Reich2000,Proctor2009,Ding2010,Sun2013} with in-plane strain.  The problem is that these shift rates are not the same.  That is what we investigate here by density functional theory (DFT) calculations \cite{DFT1,DFT2} of graphite under non-hydrostatic conditions. 

Huang \textit{et al.} gave the dynamical equations of the graphene optical phonon modes E$_{2g}$ as \cite{Huang2009,Ganesan1970,DeWolf1996}
\begin{equation}
\sum_{\beta} K_{\alpha\beta}u_{\beta}=\omega^2u_a
\label{dyn}
\end{equation}
where \textbf{u}=($u_1$,$u_2$) is the relative displacement of the two carbon atoms in the unit cell, $\omega$ is the phonon frequency, and \textit{K} is the force constant tensor, which can be expanded in powers of strain as
\begin{equation}
K_{\alpha\beta}=K_{\alpha\beta}^0+\sum_{lm}K_{\alpha\beta lm}^{\varepsilon}\varepsilon_{lm}
\label{e3}
\end{equation}

$K_{\alpha\beta lm}^{\varepsilon}$ has only two independent elements because of the hexagonal lattice, so Eqs.~\ref{dyn} and~\ref{e3} may also be written as
\begin{widetext}
\begin{equation}
\begin{pmatrix} \omega_0^2+A\varepsilon_{xx}+B\varepsilon_{yy} & 0 \\ 0 & \omega_0^2+B\varepsilon_{xx}+A\varepsilon_{yy} \end{pmatrix} \begin{pmatrix} u1 \\ u2 \end{pmatrix} = \omega^2 \begin{pmatrix} u1 \\ u2 \end{pmatrix}
\label{m1}
\end{equation}
\end{widetext}
where A and B are the two independent elements of $K_{\alpha\beta lm}^{\varepsilon}$ and $\omega_0$ is the unperturbed frequency. \textcolor{blue}{For small shifts, $\omega^2-\omega_0^2\approx 2\omega_0(\omega-\omega_0)$, Thomsen \textit{et al.} expressed the solution to the secular equation of Eq.~\ref{m1} with the Gr\"{u}neisen parameter and shear deformation potential (\textit{SDP}) as \cite{Thomsen2002}
\begin{equation}
\frac{\Delta \omega}{\omega_0}=\gamma^0(\varepsilon_{xx}+\varepsilon_{yy})\pm\frac{1}{2}SDP(\varepsilon_{xx}-\varepsilon_{yy})
\label{e2}
\end{equation}
where $\gamma^0=(A+B)/4\omega_0^2$ (the superscript 0 is used to distinguish from the refined $\gamma$ presented later in this paper) and $SDP=(B-A)/2\omega_0^2$.} 

Eq.~\ref{e2} makes explicit the two-dimensional nature of the analysis.  The Grueneisen parameter $\gamma$ and the $SDP$ are the two key parameters and a number of experimental \cite{Mohiuddin2009,Huang2009,Ni2008,Proctor2009,Soldatov2012} and theoretical \cite{Mohiuddin2009,Mohr2009,Thomsen2002} papers reported work on graphene under strain to define their accurate values. The results are shown as $\gamma^0$ and $SDP$ in TABLE~\ref{t1}. It is worth noticing that Ghandour \textit{et al.} pointed out that the transverse strain $\varepsilon_T$=0 rather than $\varepsilon_T$=-$\nu\varepsilon_L$, where $\nu$ is the in-plane Poisson's ratio and $\varepsilon_L$ is the longitudinal strain, in the case that uniaxial strain is applied by flexure of a beam to which a graphene flake adhered \cite{Ghandour2013}.
\begin{table*}
\caption{The Gr\"{u}neisen parameter and $SDP$ for $E_{2g}^{(2)}$ obtained from various experiments and calculations of graphene and graphite are listed. The values in square brackets are the corrections by Ghandour \textit{et al.} \cite{Ghandour2013}. \textcolor{blue}{$\gamma^0$(Eq.~\ref{e4}) is the value calculated for each case from Eq.~\ref{e4}, where the values of $\gamma$ and $\gamma\prime$ are the same for all the cases and the ratio of out-of-plane strain to in-plane is obtained with the approximation that graphene and graphite have the same elastic constants.} \label{t1}}
\begin{ruledtabular}
\begin{tabular}{l*{5}{c}r}
Experiments (graphene)    &$\gamma^0$ &SDP &$\gamma^0$ (Eq.~\ref{e4}) & $\gamma$ &$\gamma\prime$ &$\epsilon_{zz}/(\epsilon_{xx}+\epsilon_{yy})$ \\
\hline
uniaxial strain(beam flexure)  &0.69, \cite{Huang2009}[0.58] &0.38, \cite{Huang2009}[0.435] &1.90 &1.90 &-0.0131 &0 \\
uniaxial strain(beam flexure)  &1.99, \cite{Mohiuddin2009}[1.34] &0.99, \cite{Mohiuddin2009}[1.31] &1.90 & & &0 \\
uniaxial strain(substrate stretch)  &1.5, \cite{Ni2008} & &1.90 & & &0 \\
hydrostatic pressure  &1.99, \cite{Proctor2009} & &1.48 & & &32.25 \\
hydrostatic pressure  &2.3, \cite{Soldatov2012} & &1.48 & & &32.25 \\
\\
Calculations (graphene) \\
\hline
uniaxial strain (in-plane)  &1.87, \cite{Mohiuddin2009} &0.92, \cite{Mohiuddin2009} &1.90 & & &0 \\
uniaxial strain (in-plane)  &1.83, \cite{Mohr2009} &1.18, \cite{Mohr2009} &1.90 & & &0 \\
biaxial strain (in-plane)  &1.8, \cite{Mohiuddin2009} & &1.90 & & &0 \\
hydrostatic pressure  &2.0, \cite{Thomsen2002} & &1.48 & & &32.25 \\
shear strain  & &0.66, \cite{Thomsen2002} \\ 
\\
Experiments (graphite) \\
\hline
hyrdrostatic pressure  &1.59, \cite{Hanfland1989,Proctor2009} & &1.48 & & &32.25 \\
\end{tabular}
\end{ruledtabular}
\end{table*}

For graphite, when two adjacent graphene layers are considered, we can simply make two copies of Eq.~\ref{m1}, as 
\begin{widetext}
\begin{equation}
\begin{pmatrix} \omega_0^2+A\varepsilon_{xx}+B\varepsilon_{yy} & 0 & C & 0 \\ 0 & \omega_0^2+B\varepsilon_{xx}+A\varepsilon_{yy} & 0 & C \\ C & 0 &  \omega_0^2+A\varepsilon_{xx}+B\varepsilon_{yy} & 0 \\ 0 & C & 0 & \omega_0^2+B\varepsilon_{xx}+A\varepsilon_{yy} \end{pmatrix} \begin{pmatrix} u1 \\ u2 \\ u3 \\ u4 \end{pmatrix} = \omega^2 \begin{pmatrix} u1 \\ u2 \\ u3 \\ u4 \end{pmatrix}
\label{m2}
\end{equation}
\end{widetext}
where C is added to account for the interlayer coupling. The longitudinal modes are not coupled with the transverse modes, giving the zero elements. Eq.~\ref{e2} still applies and the weak interlayer coupling is usually neglected. Thomsen \textit{et al.}  \cite{Proctor2009} obtained the corresponding Gr\"{u}neisen parameter as 1.59 (presented in TABLE~\ref{t1}), from the experimental data of graphite under hydrostatic pressure \cite{Hanfland1989}.

We are now able to demonstrate the problem more explicitly --- if the frequency shifts of the in-plane modes are induced by in-plane strain alone, as shown in Eq.~\ref{e2}, for graphene, the shift rates of E$_{2g}$ with in-plane strain (therefore the Gr\"{u}neisen parameter) should be the same no matter how the strain is applied and for graphite, E$_{1u}$ and E$_{2g}^{(2)}$ modes should have the same shift rates as the graphene E$_{2g}$, also no matter how the strain is applied (hydrostatic or biaxial). This is against the results shown in TABLE \ref{t1}. \textcolor{blue}{It also contradicts the results from previous studies} on graphite that E$_{1u}$ shifts faster than E$_{2g}^{(2)}$ under hydrostatic pressure \cite{Hanfland1989,Perez2014,Cousins2003}.  

The different shift rates between \textcolor{blue}{E$_{1u}$ and E$_{2g}$, or E$_{2g}$} from various experiments could be a consequence of the compression of the $\pi$-electrons changing the in-plane bonds. \textcolor{blue}{To describe this effect, we suggest refining Eq.~\ref{m2}, making a phenomenological extension as
\begin{widetext}
\begin{equation}
\begin{split}
\begin{pmatrix} \omega_0^2+A\varepsilon_{xx}+B\varepsilon_{yy}+D\varepsilon_{zz} & 0 & C+E\varepsilon_{zz} & 0 \\ 0 & \omega_0^2+B\varepsilon_{xx}+A\varepsilon_{yy}+D\varepsilon_{zz} & 0 & C+E\varepsilon_{zz} \\ C+E\varepsilon_{zz} & 0 &  \omega_0^2+A\varepsilon_{xx}+B\varepsilon_{yy}+D\varepsilon_{zz} & 0 \\ 0 & C+E\varepsilon_{zz} & 0 & \omega_0^2+B\varepsilon_{xx}+A\varepsilon_{yy}+D\varepsilon_{zz} \end{pmatrix} \\ 
\begin{pmatrix} u1 \\ u2 \\ u3 \\ u4 \end{pmatrix} = \omega^2 \begin{pmatrix} u1 \\ u2 \\ u3 \\ u4 \end{pmatrix}
\label{m3}
\end{split}
\end{equation}
\end{widetext}
where D and E are the additional two independent parameters arising from the new degree of freedom along c-axis, D accounting for the compression of the $\pi$-electrons into the $sp^2$ network and E for the coupling between layers. These are clearer in the solutions to the secular equation of Eq.~\ref{m3} as
\begin{align*}
\omega_{(1)}^2 &= \omega_0^2(E_{2g}^{(2)})+(A+B)\times\varepsilon_{in-plane}+(D+E)\times\varepsilon_{zz} \\
\omega_{(2)}^2 &= \omega_0^2(E_{2g}^{(2)})+(A+B)\times\varepsilon_{in-plane}+(D+E)\times\varepsilon_{zz} \\
\omega_{(3)}^2 &= \omega_0^2(E_{1u})+(A+B)\times\varepsilon_{in-plane}+(D-E)\times\varepsilon_{zz} \\
\omega_{(4)}^2 &= \omega_0^2(E_{1u})+(A+B)\times\varepsilon_{in-plane}+(D-E)\times\varepsilon_{zz}
\end{align*}
where $\varepsilon_{xx}$ is equated to $\varepsilon_{yy}$ for the following three cases in this paper and $C$ accounts for the difference of the frequency of the E$_{1u}$ and E$_{2g}^{(2)}$ modes of unstrained graphite. The solutions lead to a new parameter $\gamma^\prime$, relating out-of-plane strain to its contribution to the shift of the in-plane phonon frequencies, added to Eq.~\ref{e2} as
\begin{flalign}
\begin{split}
&\frac{\Delta \omega}{\omega_0}=-\gamma(\varepsilon_{xx}+\varepsilon_{yy})\mp\frac{1}{2}SDP(\varepsilon_{xx}-\varepsilon_{yy})-\gamma^\prime\varepsilon_{zz} \\
&with \\
&\gamma^0=\gamma+\gamma\prime\frac{\epsilon_{zz}}{\epsilon_{xx}+\epsilon_{yy}}
\label{e4}
\end{split}
\end{flalign}
where $\gamma=(A+B)/4\omega_0^2$, $SDP=(B-A)/2\omega_0^2$, $\gamma_{E_{1u}}^\prime=(D+E)/2\omega_0^2$ and $\gamma_{E_{2g}^{(2)}}^\prime=(D-E)/2\omega_0^2$, for small shifts. In Section~\ref{s1}, we model uniaxial strain and uniaxial stress along the out-of-plane c-axis, and hydrostatic pressure on graphite, to quantify all the parameters and then explain the different shifts of the E$_{1u}$, E$_{2g}^{(2)}$ modes of graphite and the E$_{2g}$ of graphene under hydrostatic pressure.} 

\section{Methods}
Graphite was studied at 0 K using DFT \cite{DFT1,DFT2} as implemented in the Vienna Ab initio Simulation Package (VASP) \cite{VASP}. The exchange-correlation effects were treated within the generalised gradient approximation (GGA) as parameterized by Perdew, Burke and Ernzerhof \cite{GGA} and the projector augmented-wave method pseudopotentials \cite{PP} for carbon were used. To reach highly accurate results, we used 900 eV plane-wave cut-off energy, and the reciprocal unit cell was sampled with 18x18x9 k-mesh. Van der Waals (vdW) effects were included using the Grimme method \cite{vdW} as implemented in the VASP code. The elastic properties were evaluated using the stress-strain method \cite{cij}. The vibrational frequencies at the Brillouin zone centre, the $\Gamma$ point, were calculated using the 2x2x2 supercell employing the finite displacement method as implemented in the Phonopy code \cite{phonon}.

\section{Results and Discussion}
\label{s1}
\subsection{Geometry}
First of all, we obtain the optimized geometry for unstrained graphite, as the in-plane bond length of $a=1.42 \mbox{\AA}$. and the interlayer distance of $c=3.20 \mbox{\AA}$. The errors relative to the experimental values \cite{Hanfland1989} are 0.06$\%$ and 4.6$\%$. The vdW add-on is included, nevertheless the interlayer interaction is not so well-described as the in-plane covalent bonding. The LDA calculation (without vdW) usually gives a better agreement to the experimental value of the interlayer distance, however this is considered to be a coincidence because LDA is a local approximation which overestimates bonding. To minimize the effects of calculating vdW inaccurately , we study the bond anharmonicity under compressive strain, where the vdW attractive potential plays only a small role compared to the dominant repulsion. The error in the value of interlayer distance would not affect the phonon frequency shift rates with compressive strain as much as it would under tensile strain. 

\subsection{Hydrostatic compression}
\begin{figure}
\includegraphics[width=0.9\linewidth]{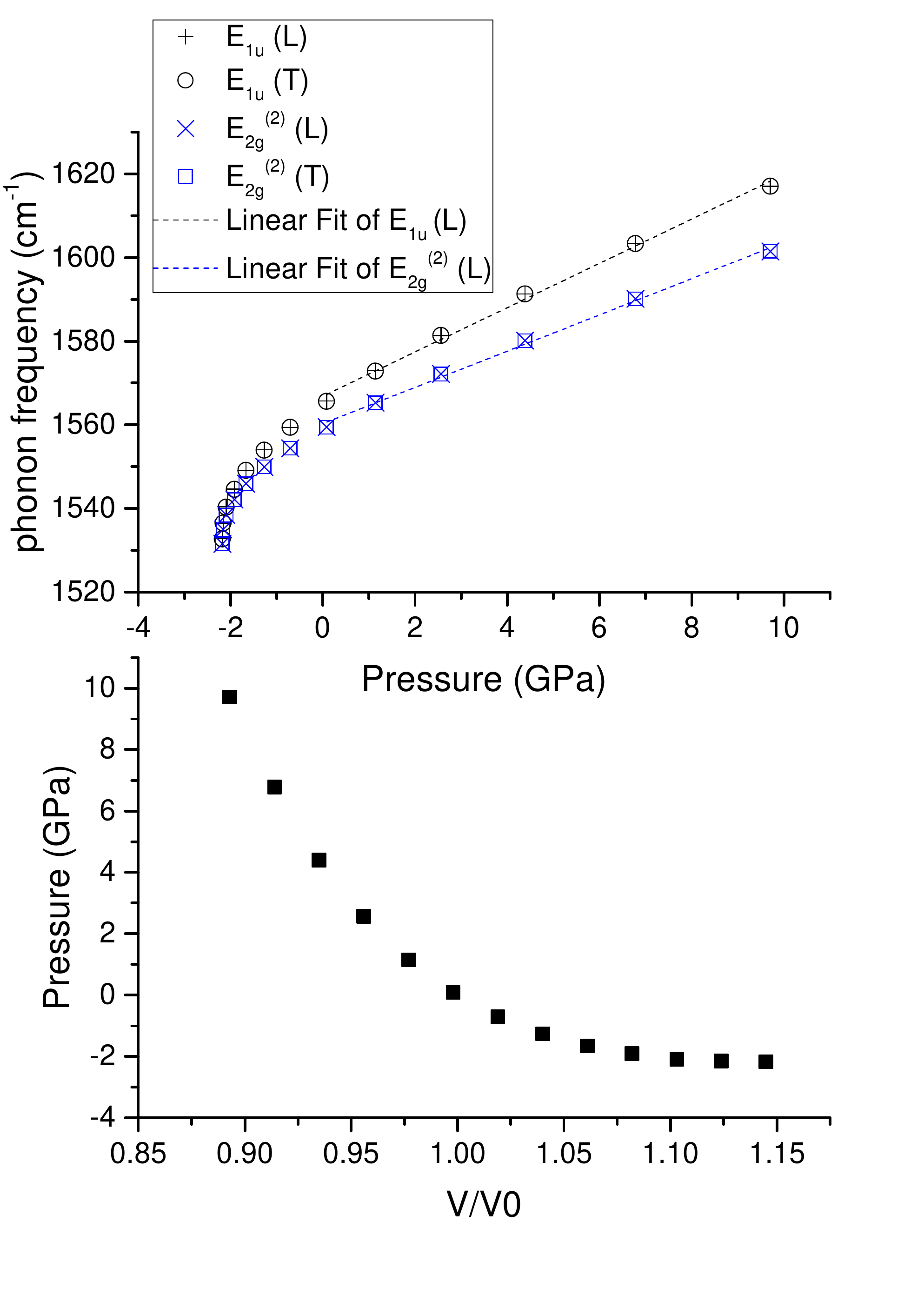}%
\caption{(colour online) [Modelling] Graphite under hydrostatic pressure. The frequencies of the graphite E$_{1u}$ and E$_{2g}^{(2)}$ are plotted against pressure. The pressures are plotted against the unit cell volumes at which they are calculated. $V_0$ is the unit cell volume of unstrained graphite. The linear fit of the phonon frequencies at compressive pressure up to 10 GPa is presented as dashed lines, black for E$_{1u}$ and blue for E$_{2g}^{(2)}$.}\label{f1}
\end{figure}

We then model hydrostatic pressure on graphite by setting a smaller unit cell volume than the unstrained, optimizing the geometry at that certain volume, and calculating the corresponding $sp^2$ bond length, interlayer distance, pressure and phonon frequencies. The frequencies of the E$_{1u}$ and E$_{2g}^{(2)}$ modes of unstrained graphite are 1565.2 and 1559.1 cm$^{-1}$, respectively. The errors relative to the experiments are 1.4$\%$ and 1.3$\%$ \cite{Hanfland1989,Nemanich1977}. We assume that they are linked to the vdW attractive term and so they would not affect the shift rates with compressive strain. Phonon frequencies are plotted against pressure in Fig.~\ref{f1}, as is the standard for presenting experimental data. And the pressure, now as a calculation output, is plotted against the input here --- the unit cell volume. (L) and (T) refer to two orthogonal in-plane vibrations, longitudinal and transverse. The frequency difference between these two under hydrostatic condition is less than 0.4 cm$^{-1}$ for both E$_{1u}$ and E$_{2g}^{(2)}$ and the shift rates of (L) and (T) with pressure are the same in the case of the E$_{1u}$ and E$_{2g}^{(2)}$. Therefore, here and in the following calculation, we treat the differnce between longitudinal and transverse modes as computational error and will study the longitudinal modes alone as a representative. Linear least square fits give the shift rates with compressive pressure up to 10 GPa at 5.3 and 4.3 cm$^{-1}$GPa$^{-1}$ for E$_{1u}$ and E$_{2g}^{(2)}$ modes, respectively. No experimental data for E$_{1u}$ exists and the shift rates for E$_{2g}^{(2)}$ were \cite{Hanfland1989,Liu1990,Sandler2003} 4.1--4.6. In the previous theoretical work, Cousins \textit{et al.} obtained 4.74 and 4.67 cm$^{-1}$GPa$^{-1}$ for E$_{1u}$ and E$_{2g}^{(2)}$ modes \cite{Cousins2003}, while Abbasi-P\'{e}rez \textit{et al.} \cite{Perez2014} got 5.0 and 4.3 cm$^{-1}$GPa$^{-1}$. To summarize, the calculation results are reliable and reasonable, with the shift rates with pressure comparable to previous work, with clear sublinearity of the frequency shift due to the pressure dependence of the elastic constant $C_{33}$, and with the two in-plane modes degenerate when the graphene layers are pulled apart. However, the problems are again the different shift rates for E$_{1u}$ and E$_{2g}^{(2)}$ with pressure, and the behaviour of the frequency starting off vertically upwards with pressure (see FIG.~\ref{f1}). The latter point implies that the pressure (force) may be inaccurately calculated under tensile stress (pressure remains at about -2GPa when the unit cell volume keeps increasing), where vdW plays an important part. We will resolve the former point and we avoid the latter point by focusing on the compressive part.

\subsection{Non-hydrostatic compression}
Modelling non-hydrostatic condition helps to investigate problems found under hydrostatic condition. The only study of graphite under non-hydrostatic condition reported so far is that of Abbasi-P\'{e}rez \textit{et al.}.  They suggested that the contribution to the shift of the in-plane phonon frequency from the out-of-plane compression is so little that it can be neglected \cite{Perez2014}. The following results do not agree with that.

\begin{figure*}
\includegraphics[width=0.9\linewidth]{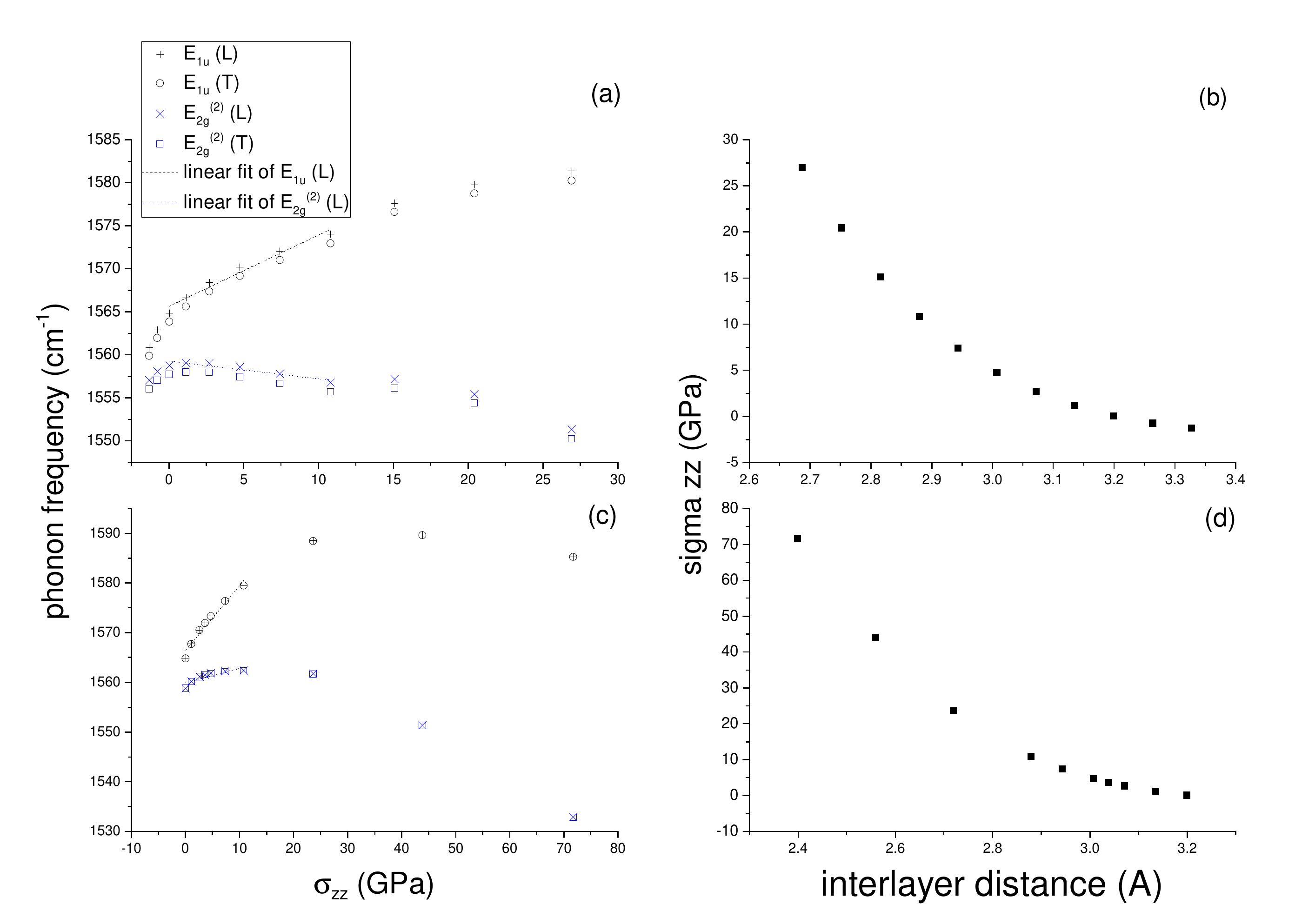}%
\caption{(colour online) [Modelling] Graphite under uniaxial strain / stress along the c-axis. The frequencies of the graphite E$_{1u}$ and E$_{2g}^{(2)}$ are plotted against out-of-plane stress in the case of uniaxial strain (a) and unaixial stress (c). The out-of-plane stresses are plotted against the interlayer distances at which they are calculated, in the case of uniaxial strain (b) and uniaxial stress (d). The linear fit of the phonon frequencies at compressive stress to about 10 GPa is presented as dashed lines, black for E$_{1u}$ and blue for E$_{2g}^{(2)}$.} \label{f2}
\end{figure*}

We model uniaxial strain along the c-axis of graphite by varying the interlayer distance while fixing the in-plane geometry. The phonon frequencies and out-of-plane stress are calculated at each interlayer distance and fixed $sp^2$ bond length. FIG.~\ref{f2} (a) shows the shifts of the in-plane phonon frequencies against out-of-plane stress along c-axis. In FIG.~\ref{f2} (b), the stress as a calculation output, is plotted against the input --- the interlayer distance in this case. The shift rates with stress up to about 10 GPa, by least square linear fits, are 0.8 and -0.2 cm$^{-1}$GPa$^{-1}$ for E$_{1u}$ and E$_{2g}^{(2)}$, respectively. It is worth noticing that the elastic constant $C_{13}$, determining the Poisson's ratio $\nu_{zx}$, is poorly defined due to the structural anisostropy of graphite, but can be considered to be close to zero \cite{Bosak2007}. Our calculated $C_{13}$ value is -10.5 GPa, and the corresponding $\nu_{zx}$ is -0.024. Therefore, uniaxial compressive strain here induces in-plane tensile stress. The degeneracy of the two modes can be again seen in this case when the graphene sheets are pulled apart. The problem is that the E$_{1u}$ and E$_{2g}^{(2)}$ modes shift with opposite signs.

\begin{figure}
\includegraphics[width=0.9\linewidth]{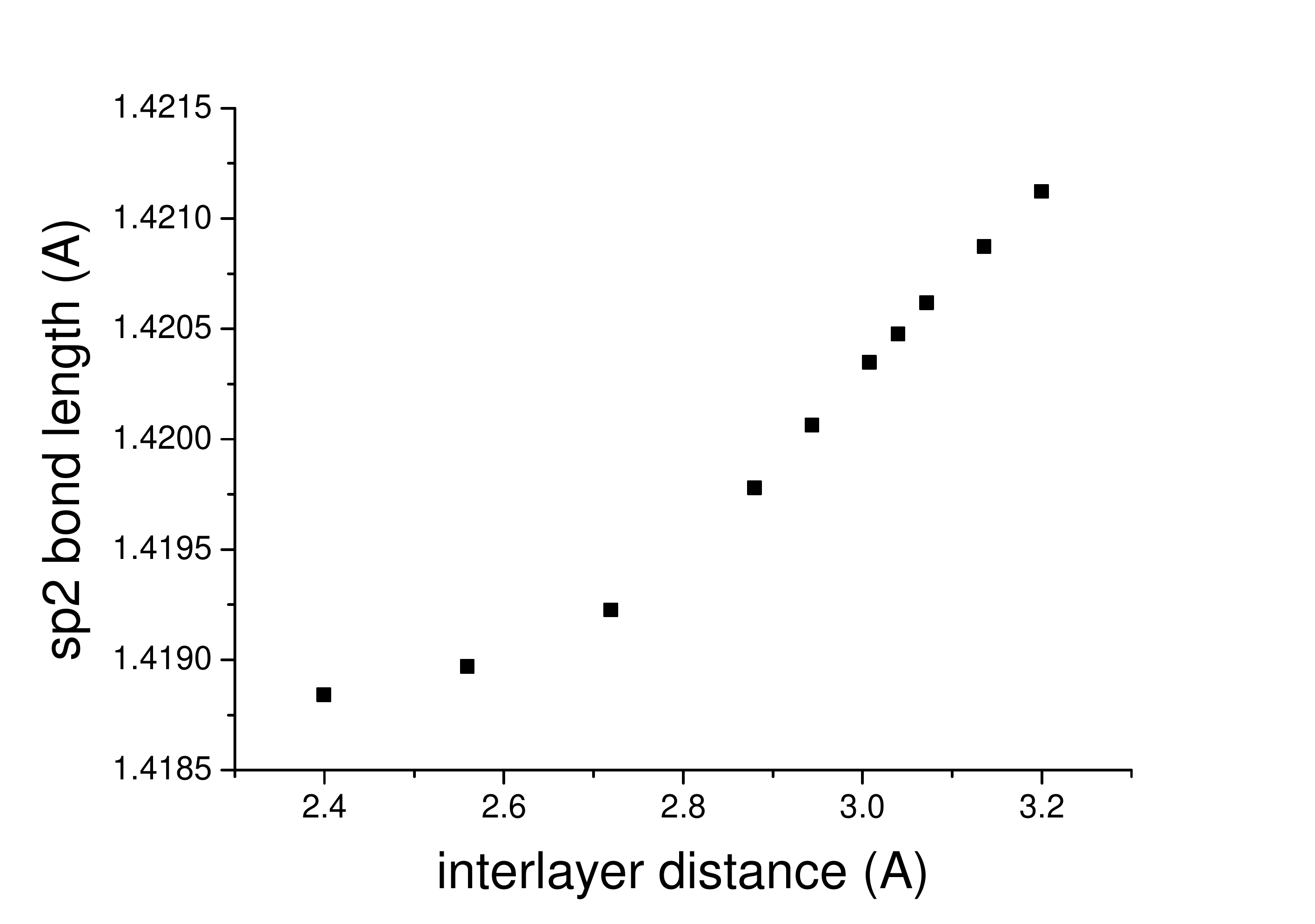}%
\caption{(colour online) [Modelling] Graphite under uniaxial stress along c-axis. The $sp^2$ bond length is plotted against the interlayer distance, at which it is calculated.} \label{f7}
\end{figure}

Next we consider uniaxial stress on graphite along the c-axis, by varying the interlayer distance and optimizing the in-plane geometry at each interlayer distance. The phonon frequencies, the out-of-plane stress and the $sp^2$ bond length are calculated at each interlayer distance. The effect of the negative Poisson's ratio can now be clearly illustrated in FIG.~\ref{f7} as the in-plane bond is also compressed as we compress along the c-axis. The amount, however, is tiny. FIG.~\ref{f2} (c) presents the in-plane phonon frequency against out-of-plane stress and again the output stress is plotted against the input interlayer distance in FIG.~\ref{f2} (d). The shift rates with stress up to 10 GPa in this case are 1.3 and 0.3 cm$^{-1}$GPa$^{-1}$ for E$_{1u}$ and E$_{2g}^{(2)}$, respectively. The shift rate with uniaxial stress for the E$_{1u}$ is about a quarter of the shift rate under hydrostatic stress; this is large enough to be significant. 

It is usual to present frequency against stress, because stress is considered as input in experiments. However, the atomic positions (strain) determine properties such as phonon frequency, and it is therefore helpful to plot the frequencies against strain. To be specific, for graphite, the shifts of the frequencies of the in-plane modes E$_{1u}$ and E$_{2g}^{(2)}$ are considered to be induced by in-plane strain. So we plot the phonon frequencies against in-plane strain under hydrostatic and non-hydrostatic conditions and then obtain the corresponding Gr\"{u}neisen parameters $\gamma_{E_{1u}}$ and $\gamma_{E_{2g}^{(2)}}$ for the E$_{1u}$ and E$_{2g}^{(2)}$ modes, respectively, according to Eq.~\ref{e2}. Values for the two modes should be the same from the two dimensional nature of Eq.~\ref{dyn}--\ref{e2}.

In the case of uniaxial strain, the phonon frequencies for both modes shift at fixed in-plane geometry. The Gr\"{u}neisen parameters are hence $\gamma_{E_{1u}}$=$\infty$ and $\gamma_{E_{2g}^{(2)}}$=$\infty$, according to Eq.~\ref{e2}.

\begin{figure}
\includegraphics[width=0.9\linewidth]{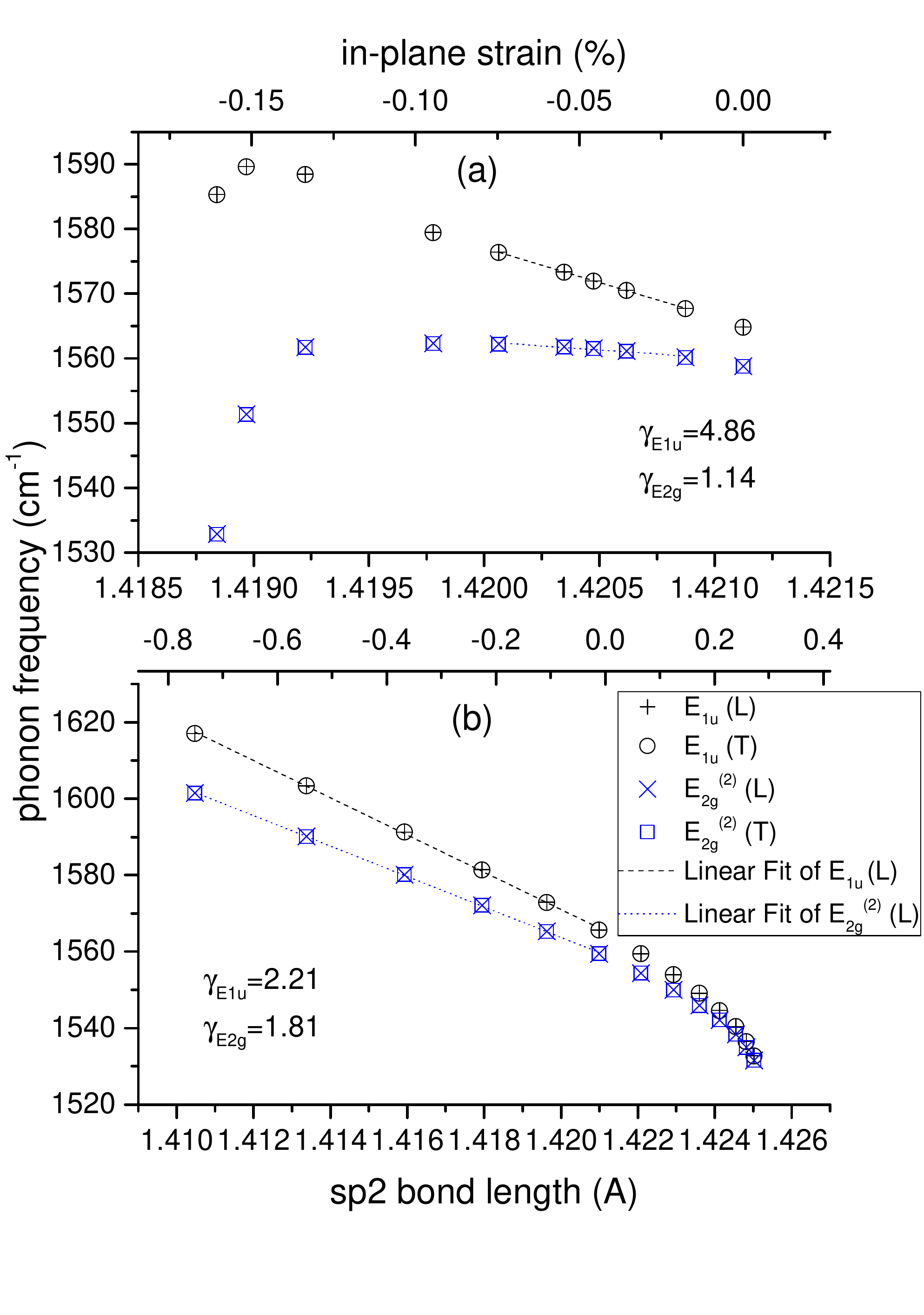}%
\caption{(colour online) [Modelling] Graphite under uniaxial stress along c-axis (a) / hydrostatic pressure (b). The data in (a) is the same as in FIG. 2 (c) and the data in (b) is the same as in FIG. 1. Here the frequencies of the graphite E$_{1u}$ and E$_{2g}^{(2)}$ are plotted against in-plane $sp^2$ bond length, which is calculated at each interlayer distance (a) / unit cell volume (b). The top axis of in-plane strain is converted from the $sp^2$ bond length by $\varepsilon = (a-a_0)/a_0 \times 100 \%$. The linear fit of the phonon frequencies at compressive stress to about 10 GPa is presented as dashed lines, black for E$_{1u}$ and blue for E$_{2g}^{(2)}$. The Gr\"{u}neisen parameters for both modes, obtained from the linear fit, are presented in each case for comparison to the values in TABLE~\ref{t1}.} \label{f4}
\end{figure}

In the case of uniaxial stress, in FIG.~\ref{f4} (a) we plot the in-plane phonon frequencies against the $sp^2$ bond length, which is calculated by optimizing the in-plane geometry at each interlayer distance. The top axis of in-plane strain is converted from the $sp^2$ bond length by $\varepsilon = (a-a_0)/a_0 \times 100 \%$, where $\varepsilon$ is the in-plane strain, $a$ is the $sp^2$ bond length and $a_0$ is the $sp^2$ bond length of unstrained graphite. This is the same data as in FIG.~\ref{f2}. We apply a linear fit to the data points under compression up to about 10 GPa (the same as in FIG.~\ref{f2}), and obtain the shift rates $\partial\omega_{E_{1u}}/\partial\varepsilon$=-152.00 cm$^{-1}$/\% and $\partial\omega_{E_{2g}^{(2)}}/\partial\varepsilon$=-35.50 cm$^{-1}$/\%, which correspond to $\gamma_{E_{1u}}$=4.86 and $\gamma_{E_{2g}^{(2)}}$=1.14.  

In the case of hydrostatic pressure, in FIG.~\ref{f4} (b) we plot in-plane phonon frequency against the $sp^2$ bond length, which is calculated by the geometry optimization at each unit cell volume. The top axis of in-plane strain is converted in the same way as before. The data is the same as in FIG.~\ref{f1}. We apply a linear fit to the data points under compression up to about 10 GPa (the same as in FIG.~\ref{f1}) and get the shift rates $\partial\omega_{E_{1u}}/\partial\varepsilon$=-69.20 cm$^{-1}$/\% and $\partial\omega_{E_{2g}^{(2)}}/\partial\varepsilon$=-56.59 cm$^{-1}$/\%, corresponding to $\gamma_{E_{1u}}$=2.21 and $\gamma_{E_{2g}^{(2)}}$=1.81.  

We present the Gr\"{u}neisen parameters obtained in the case of uniaxial stress and hydrostatic pressure in FIG.~\ref{f4} and compare them to that of uniaxial strain and those reported in TABLE~\ref{t1}. The values for the E$_{1u}$ and E$_{2g}^{(2)}$ modes obtained under hydrostatic condition are in good agreement with the most values of the E$_{2g}$ mode of graphene presented in TABLE~\ref{t1}. Similar agreement has been reported in the literature; this is the quantitative reason why the in-plane phonon frequency shifts are considered as induced by in-plane strain alone \cite{Thomsen2002,Mohiuddin2009,Reich2000,Proctor2009,Ding2010}. However, the difference between the values of the E$_{1u}$ and E$_{2g}^{(2)}$ modes \textcolor{blue}{increases under uniaxial strain when the ratio of out-of-plane strain to in-plane is considerably larger than it is under hydrostatic pressure} and neither of the values agrees with the E$_{2g}$ of graphene. Under uniaxial strain, where there is only out-of-plane strain, the difference of the frequency shifts can be considered as infinity. The out-of-plane strain is responsible for the difference of the Gr\"{u}neisen parameters for the E$_{1u}$ and E$_{2g}^{(2)}$ modes. Regarding the amount of the difference, to include the out-of-plane strain contribution to the in-plane phonon frequency is desirable in the case of hydrostatic pressure and definitely necessary in the cases of uniaxial strain and stress. 

To quantify this contribution, \textcolor{blue}{we refined Eq.~\ref{m2} as Eq.~\ref{m3}. The solutions to the secular equation of Eq.~\ref{m3} led to the new parameter $\gamma^\prime$, as shown in Eq.~\ref{e4}, for small shifts. Alternatively, we can rewrite Eq.~\ref{e4} with a full hydrostatic term ($\varepsilon_{xx}+\varepsilon_{yy}+\varepsilon_{zz}$) as}
\begin{equation}
\frac{\Delta \omega}{\omega_0}=-\gamma(\varepsilon_{xx}+\varepsilon_{yy}+\varepsilon_{zz})\mp\frac{1}{2}SDP(\varepsilon_{xx}-\varepsilon_{yy})-(\gamma^\prime-\gamma)\varepsilon_{zz}
\end{equation}
We believe this out-of-plane contribution is mostly related to the compression of the $\pi$ electrons, which is beyond the picture of the force constant model.  

\begin{figure}
\includegraphics[width=0.8\linewidth]{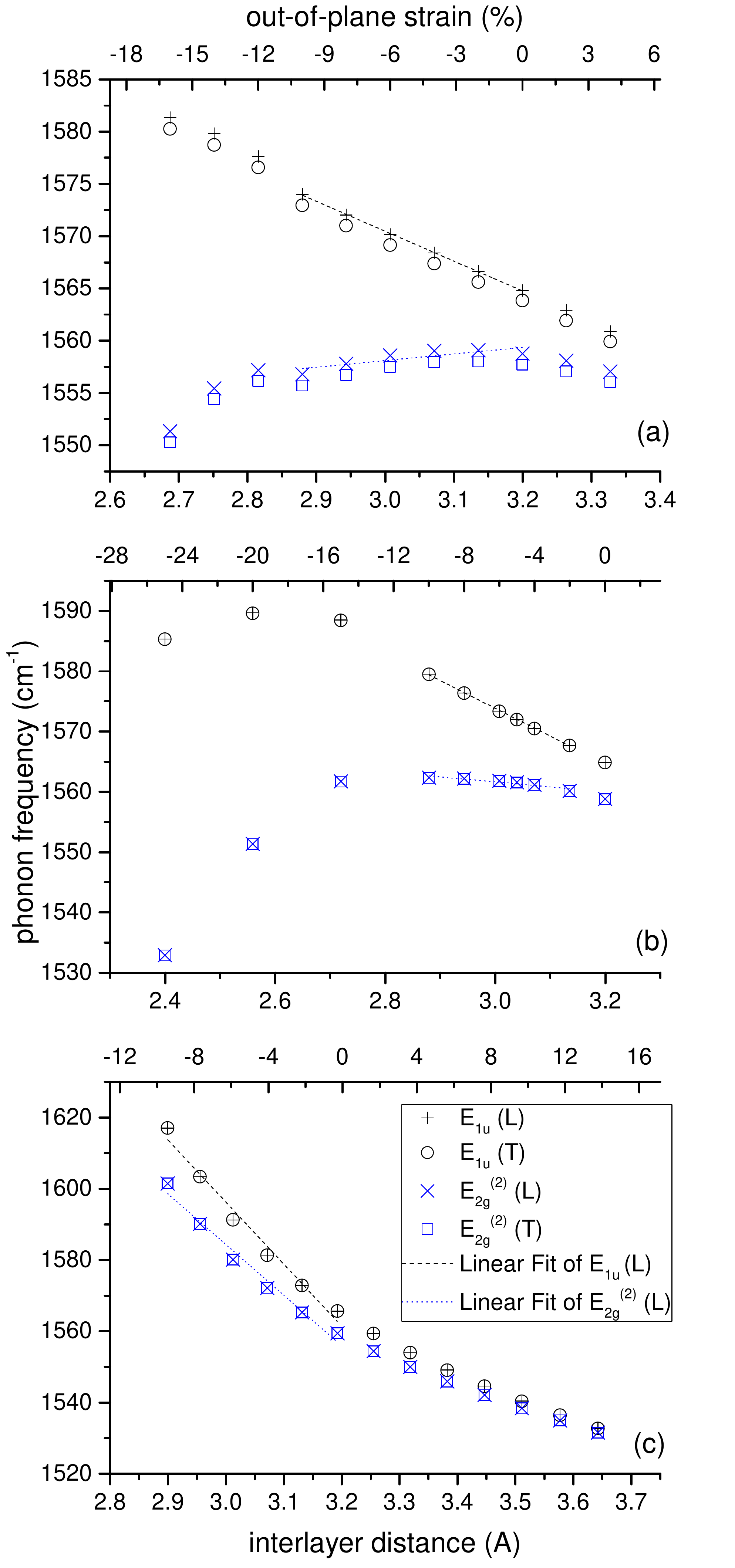}%
\caption{(colour online) [Modelling] Graphite under uniaxial strain (a) / uniaxial stress (b) / hydrostatic pressure (c). The frequencies of the in-plane modes of the graphite E$_{1u}$ and E$_{2g}^{(2)}$ are plotted against interlayer distance, which is the calculation input in the case of (a)\&(b) and is calculated at each unit cell volume in the case of (c). The top axis of out-of-plane strain is converted from interlayer distance by $\varepsilon = (a_{33}-a_{33_0}/a_{33_0}) \times 100 \%$.  The linear fit of the phonon frequencies at compressive stress to about 10 GPa is presented as dashed lines, black for E$_{1u}$ and blue for E$_{2g}^{(2)}$.} \label{f6}
\end{figure}

Let us now return to Eq.~\ref{e4}. In the case of uniaxial strain, where the shifts of the frequencies are entirely from the out-of-plane strain, we plot the in-plane phonon frequencies against the interlayer distance --- the calculation input, in FIG.~\ref{f6} (a) and fit the data under compression up to about 10 GPa (the same as in FIG.~\ref{f2} (a)). The top axis of out-of-plane strain is converted from interlayer distance by $\varepsilon = (a_{33}-a_{33_0}/a_{33_0}) \times 100 \%$, where $\varepsilon_o$ is the out-of-plane strain, $a_{33}$ is the interlayer distance and $a_{33_0}$ is the value of unstrained graphite. The shift rates for the E$_{1u}$ and E$_{2g}^{(2)}$ modes are $\partial\omega_{E_{1u}}/\partial\varepsilon$=-0.915 cm$^{-1}$/\% and $\partial\omega_{E_{2g}^{(2)}}/\partial\varepsilon$=0.204 cm$^{-1}$/\%, corresponding to $\gamma^\prime_{E_{1u}}$=0.0585 and $\gamma^\prime_{E_{2g}^{(2)}}$=-0.0131, according to Eq.~\ref{e4}. They are small, but non-negligible as the out-of-plane strain is about 30 times larger than the in-plane strain under hydrostatic condition (from the anisotropy of graphite) and can be even larger under non-hydrostatic conditions. It is worth noticing that the in-plane phonon frequency cannot be considered as an indicator of the in-plane bond stiffness in this case as the E$_{1u}$ and E$_{2g}^{(2)}$ modes, both representing the in-plane bond stiffness, shift with opposite signs under out-of-plane compressive strain. Now we have quantified the out-of-plane strain contribution by $\gamma^\prime$, which is responsible for the separating of the E$_{1u}$ and E$_{2g}^{(2)}$ modes and then the in-plane $\gamma$ can be the same in various conditions for the two modes (and the E$_{2g}$ of graphene) as it should be from its definition.

Finally, we calculate the refined value for the in-plane $\gamma$ from Eq.~\ref{e4}, by $\gamma^\prime_{E_{1u}}$ and $\gamma^\prime_{E_{2g}^{(2)}}$ obtained under uniaxial strain. For both modes, in the case of uniaxial stress and hydrostatic pressure, we calculate the contribution to the shifts of the frequencies from out-of-plane strain by $\gamma^\prime$ and attribute the rest to the in-plane strain. And from that we obtain the refined in-plane $\gamma$.

We plot the in-plane phonon frequencies against interlayer distance, which is the calculation input under uniaxial stress in FIG.~\ref{f6} (b) and calculated by the geometry optimization at each unit cell volume under hydrostatic pressure in FIG.~\ref{f6} (c). The top axis of the out-of-plane strain is converted from the interlayer distance in the way as mentioned above. The data is the same as in FIG.~\ref{f2} (c) and \ref{f1}, for uniaxial stress and hydrostatic pressure, respectively. We obtain the shift rates for the two modes at the same pressure range as before. Under uniaxial stress, we obtain $\gamma_{E_{1u}}$=1.84 and $\gamma_{E_{2g}^{(2)}}$=2.01 from the results shown in FIG.~\ref{f4} (a) and \ref{f6} (b). Under hydrostatic pressure, the values are $\gamma_{E_{1u}}$=1.85 and $\gamma_{E_{2g}^{(2)}}$=1.90, from the results shown in FIG.~\ref{f4} (b) and \ref{f6} (c).

\textcolor{blue}{\subsection{Summary of the results and their applications}
The E$_{2g}^{(2)}$ and E$_{1u}$ will be separately discussed so the subscript of the Gr\"{u}neisen parameter is removed in this paragraph. For E$_{2g}^{(2)}$, the uniaxial strain modelling gave $\gamma\prime=-0.0131$. Applying this $\gamma\prime$ to uniaxial stress modelling we obtained $\gamma=2.01$ and to hydrostatic pressure modelling we obtained $\gamma=1.90$. The discrepancy is due to non-linear relationship shown in FIG.~\ref{f6} (a). We choose to use $\gamma=1.90$ because the data of hydrostatic pressure modelling was validated by comparing to experiments. Now Eq.~\ref{e4} can be used to give the value of $\gamma^0$ for each case in TABLE~\ref{t1} with the approximation that graphene and graphite have the same elastic constants. The comparison is presented in the table. No firm explanation can be made of the current results of graphene as the elastic constants (especially $C_{13}$ and $C_{33}$) of graphene cannot be accurately obtained and also due to the errors in these experiments, but it is clear that graphite has a smaller $\gamma^0$ than graphene because of $\gamma\prime$. For the studies of graphite under uniaxial compression along c-axis \cite{Perez2014,Taravillo2014}, $\epsilon_{zz}\gg(\epsilon_{xx}+\epsilon_{yy})$, the contribution of $\gamma\prime$ becomes much more significant. Refinement can be done to include the mentioned non-linear effect but should await further work --- either initial experimental evidence or, theoretically, the evolution of the $\pi$-electrons can be visualised under compression to provide a clearer picture of the phenomenon to be quantified. For E$_{1u}$, the uniaxial strain modelling gave $\gamma\prime$=0.0585. Applying this $\gamma\prime$ to uniaxial stress modelling we obtained $\gamma=1.84$ and to hydrostatic pressure modelling we obtained $\gamma=1.85$ --- excellent agreement achieved. The Gr\"{u}neisen parameter of E$_{1u}$ was considered the same as E$_{2g}^{(2)}$ and indeed they are close. But the $\gamma\prime$ of opposite sign for these two modes makes them distinguishable under hydrostatic pressure and further under uniaxial compression.}

Other analysis can also be done to study the phase transition induced by the change of interlayer distance, which has clear signs in the presented results, namely the significant drop of phonon frequency in FIG.~\ref{f2} (b), \ref{f4} (a) and \ref{f6} (b) under large compression.

\section{Conclusion}
We model uniaxial strain, stress along c-axis and hydrostatic pressure on graphite and calculate the vibrational frequencies of the in-plane modes derived from the graphene E$_{2g}$ mode. The shifts of the frequencies come from both in-plane and out-of-plane compression. We quantify the contribution from out-of-plane strain by \textcolor{blue}{new parameters $\gamma\prime_{E_{2g}}$=-0.0131 and $\gamma\prime_{E_{1u}}$=0.0585,} and therefore refine the existing values of the Gr\"{u}neisen parameter $\gamma$ as \textcolor{blue}{$\gamma_{E_{2g}}$=1.90 and $\gamma_{E_{1u}}$=1.85}. This contribution is responsible for the separating shifts of the E$_{1u}$, E$_{2g}^{(2)}$ modes of graphite and the E$_{2g}$ of graphene under hydrostatic pressure and therefore non-negligible, against previous conclusion. It can be significant under non-hydrostatic condition. A more reliable value of the in-plane Gr\"{u}neisen parameter is useful for strain calibration in various applications and can be further refined by studying the $\pi$-electron behaviour.

\begin{acknowledgments}
YWS thanks the Chinese Scholarship Council (CSC) for financial support. \textcolor{blue}{The authors acknowledge referees' constructive suggestions.}
\end{acknowledgments}

\bibliography{apstemplate}

\end{document}